\documentclass[]{aastex631}
\usepackage{bm}
\usepackage{color}

\newcommand{\msun}{{\rm M}_\odot}
\newcommand{\msunyr}{{\rm M}_\odot\,{\rm yr}^{-1}}

\newcommand{\cc}{{\rm cm^{-3}}}

\newcommand{\kms}{{\rm km\,s^{-1}}}

\newcommand{\vect}[1]{\mbox{\boldmath$#1$}}

\shorttitle{Complex Structure and Interchange Instability}
\shortauthors{Machida and Basu}

\begin{document}
\title{Complex Structure around a Circumstellar Disk Caused by Interchange Instability}

\correspondingauthor{Masahiro N. Machida}
\email{machida.masahiro.018@m.kyushu-u.ac.jp}

\author[0000-0002-0963-0872]{Masahiro N. Machida}
\affiliation{Department of Earth and Planetary Sciences, Faculty of Science, Kyushu University, Fukuoka 819-0395, Japan}
\affiliation{Department of Physics and Astronomy, University of Western Ontario, London, ON N6A 3K7, Canada}

\author[0000-0003-0855-350X]{Shantanu Basu}
\affiliation{Department of Physics and Astronomy, University of Western Ontario, London, ON N6A 3K7, Canada}
\affiliation{Canadian Institute for Theoretical Astrophysics, University of Toronto, 60 St. George, St., Toronto, ON M5S 3H8, Canada}

\begin{abstract}
We perform a three-dimensional nonideal magnetohydrodynamic simulation of a strongly magnetized cloud core and investigate the complex structure caused by the interchange instability. This is the first simulation that does not use a central sink cell and calculates the long term ($> 10^4$ yr) evolution even as the disk and outflow formation occur.
The magnetic field dissipates inside the disk, and magnetic flux accumulates around the edge of the disk,
leading to the occurrence of interchange instability. 
During the main accretion phase, the interchange instability occurs recurrently, disturbing the circumstellar region and forming ring, arc, and cavity structures. These are consistent with recent high-resolution observations of circumstellar regions around young protostars.
 The structures extend to $>1,000$\,au and persist for at least 30,000 yr after protostar formation, 
demonstrating the dynamic removal process of magnetic flux during star formation.
We find that the disk continues to grow even as interchange instability occurs, by accretion through channels between the outgoing cavities.
The outflow is initially weak, but becomes strong after $\sim 10^3$ yr. 
\end{abstract}

\section{Introduction}
\label{sec:intro}
Recent observations have unveiled complex fine structures around circumstellar disks during the main accretion phase \citep{tokuda14, tokuda23, tokuda24, favre20, harada23, ohashi23, shoshi24, riaz24, fielder24}. 
Some structures are quite complex, and include arcs, cavities, and spikes around the disk (see, e.g., Fig. 2 of \citealt{tokuda18}).
It has been considered that these structures are related to binary formation, fragmentation, and/or disk gravitational instability in a turbulent environment \citep{tokuda18}, in addition to external mass inflow \cite[see review by][]{pineda23}. 
However, it is difficult for some structures (such as cavities and spikes) to be explained by external mass inflow and there is insufficient mass to induce gravitational instability and fragmentation \citep{tokuda24}.
Thus, these structures have remained
largely a mystery.
It remains crucial to clarify the disk and protostar formation process during the main mass accretion phase in order to comprehensively understand star and planet formation.

Using the highest spatial resolution observations by ALMA, \citet{tokuda23, tokuda24} pointed out that these complex structures could be caused by interchange instability \citep[see also][]{fielder24, tanious24}. 
Interchange instability occurs when neighboring flux tubes interchange positions in order to reach a lower energy state. In compressible gas, the interchange instability can be related to a buoyancy instability \citep{Hughes1988}, in which a flux tube with stronger magnetic field and lower density exchanges position with another that has weaker magnetic field and greater density. In a rotating gas structure with a poloidal magnetic field, \cite{spruit95} showed that interchange instability can occur through incompressible motions of flux tubes in a region of strong gradient in the local mass-to-flux ratio, if the magnetic field is a strong source of radial support against gravity.
The arc, cavity, and spike structures were proposed to be formed through the magnetic flux leaking from the circumstellar disk due to interchange instability. 
\citet{tokuda24} also showed that, for MC27, interchange instability seems to occur recurrently and the structure is maintained for $\sim 10^5$ yr. 
In addition, \citet{tokuda23} proposed that a ring of radius $\sim3,500$\,au is caused by the magnetic flux removal process.
Several additional objects observed with ALMA have complex circumstellar structures that may be caused by interchange instability (K. Tokuda et al., in preparation), but they remain unpublished due to a lack of proper interpretation.
Theoretical research that confirms structures formed by interchange instability and emphasizes their role in the star formation process is crucial for understanding star formation and improving the interpretation of current and future observations.

The condition responsible for interchange instability can be analyzed based on the distribution of the mass-to-flux ratio.
For interchange instability to occur, the molecular cloud core must have a strong magnetic field, and a subregion needs to develop an inversion (outward gradient) of the mass-to-flux ratio \citep{spruit90, spruit95}. 
Using Zeeman measurements, \citet{crutcher99} showed that the magnetic field in prestellar cores is strong enough that $\mu\simeq2-3$ (see also \citealt{crutcher10}), where $\mu$ is the mass-to-flux ratio normalized to the critical value for collapse. 
They also showed that some prestellar cores are in a subcritical state with $\mu\le 1$. 
Recent polarization observations, using the Davis-Chandrasekhar-Fermi method, indicate that $\mu \lesssim 1$ in some molecular cloud cores \citep{eswaraiah21,Sharma22,priestley22,yen23,lin24}. 
These observations expand the scope of the already significant magnetic flux problem of star formation, which refers to the
fact that the magnetic flux in molecular cloud cores is at least several orders of magnitude greater than that in protostars \citep{nakano84}. 
To resolve this issue, magnetic flux must be removed from the central region of the collapsing core where the disk and protostar form.
\citet{nakano02} pointed out that the magnetic field dissipates in the circumstellar disk on a diffusive time scale.
On the other hand, \citet{machida20b} proposed that both magnetic dissipation and the dynamically-occurring interchange instability play a significant role in removing magnetic flux from the star and disk-forming region when the initial prestellar core has a strong magnetic field with $\mu\lesssim1$.

The complex or cavity structures caused by interchange instability in three-dimensional simulations were first presented by \citet{zhao11}. 
Subsequently, the structures created by interchange instability have been seen in many three-dimensional simulations using different numerical codes \citep{seifried11,joos12,li13,tomida15,masson16,matsumoto17,vaytet18,mignon21,tsukamoto23}. 
As described in \citet{machida20b}, interchange instability occurs when magnetic flux accumulates near the edge of the circumstellar disk. The magnetic flux is transported from the inner to the outer regions of the disk through ohmic dissipation and ambipolar diffusion.
Thus, the inclusion of nonideal MHD effects and the high spatial resolution to resolve the circumstellar disk are necessary to realistically investigate whether interchange instability occurs.
{This is because 
 interchange instability can also occur numerically when using a sink cell. This is especially true if matter is transported into a sink cell but magnetic flux is not, resulting in a mock magnetic diffusion with the mass-to-flux ratio having an outward gradient just outside the sink cell. 
 \citet{zhao11} had modeled interchange instability using an ideal MHD code in which the sink cell treatment as described above was an ad-hoc means of magnetic field dissipation.
The sink-cell method is not well suited for investigating interchange instability even with a nonideal MHD code because 
the numerically-induced inversion of mass-to-flux ratio is difficult to separate from a physical inversion due to magnetic dissipation
\citep[see][]{machida14,machida16}. 
\citet{machida20b} employed nonideal MHD, did not include a sink, and resolved the protostar with a spatial resolution of 0.01\,au. 
However, due to the high resolution, the calculation could be performed for only 500 yr after protostar formation. 
To compare the simulation with observations, it is necessary to run the calculation for $>10^4$ yr, as the main accretion phase lasts approximately $10^5$ yr. 
In this new study, we achieve a long-term calculation while resolving the protostar and using an adiabatic equation of state (EOS). We discover outcomes that reproduce observational characteristics that were proposed to be related to interchange instability \citep{tokuda24}.

\section{Model and Numerical Methods}
\label{sec:model}
This study is a continuation of the work by \citet{machida18} and \citet{machida20b}. 
The difference between this study and our previous ones is the equation of state adopted in the simulations. 
In this Letter, we present the results for a single parameter set (initial magnetic field strength and rotation rate), while the results for different parameter sets will be submitted to a main journal. 
Detailed numerical settings will also be described in the main journal.
Thus, we briefly explain our model and numerical settings in the following.

The equations [3]--[7] in \citet{machida18} are used in this study. 
The same basic equations are also used in our other studies \citep{higuchi18, higuchi19, machida20b}.
The coefficients for ohmic dissipation and ambipolar diffusion are taken from \citet{machida18} and \citet{machida20b}.
We adopt the barotropic equation of state
\begin{equation}
P = c_{s,0}^2 \rho_{\rm c} \left( \frac{\rho}{\rho_{\rm c}} \right)^{5/3},
\end{equation}
where $\rho_{\rm c}=2.0\times10^{-14}\,{\rm g}\,\cc$, corresponding to number density $n_c=5 \times10^9\,\cc$.
It differs somewhat from the equation of state used by \citet{machida18} and \citet{machida20b}.

Referring to observations in low-mass star-forming regions, we determine the parameters of the initial cloud. 
As the initial condition, we adopt a sphere with a Bonnor-Ebert (B.E.) density profile, with a central density $10^6\,\cc$ and an isothermal temperature of $10$\,K. 
To promote collapse, the density is enhanced by a factor of 1.8 \citep{machida13}. 
Note that the density enhancement factor determines the thermal stability of the core as $\alpha_0 = 0.84/f$. 
The thermal stability adopted in this study $\alpha_0 = 0.47 (= 0.84/1.8)$ is typical for prestellar cloud cores in low-mass star-forming regions \citep{Tatematsu2016}. 
The cloud's central density, $n_{\rm c,0} = 10^6,\mathrm{cm}^{-3}$, is determined based on \citet{Tokuda2020}. Combining the central density and isothermal temperature, the B.E. density profile (or its mass and radius) is uniquely determined.
The mass and radius of the Bonnor-Ebert sphere are $0.74\,\msun$ and $4.4\times10^3$\,au, respectively. 
To break the symmetry and better match observations, an $m=2$ mode and a $m=3$ mode of density perturbation are added to the initial state at 1\% and 0.1\% levels, respectively,  which is determined based on our previous works. (for details, see eq.~[14] of \citealt{machida05a}).

A uniform magnetic field $B_0 = 1.8\times10^{-4}$\,G and a rigid rotation $\Omega_0=2.3\times10^{-13}$\,s$^{-1}$ are adopted. 
The thermal, rotational, and magnetic energies normalized by the magnitude of gravitational energy are $\alpha=0.47$, $\beta=0.03$, and $\gamma=0.89$, respectively.
The mass-to-flux ratio normalized by the critical value $(2\pi G^{1/2})^{-1}$ \citep{nakano78} is $\mu=1$ for the whole cloud core and is less than unity in the inner region of the initial cloud
(see Fig.~1 of \citealt{machida18}).
It should be noted that the angular velocity is set to achieve $\beta_0 = 0.03$, which is slightly larger than the observed average value of $\beta_0 = 0.02$ \citep{Caselli2002}. The magnetic field strength is chosen to yield $\mu_0 = 1$, which falls within the observed range \citep{crutcher10}. 
Thus, the parameters used in this study are consistent with those observed in low-mass star-forming regions \citep[see also, review by ][]{tsukamoto23b}.

To calculate the evolution of the B.E. core, we use our nested grid code \citep{machida04,machida13,machida18}. Each grid has a size  of ($i, j, k$) = (128, 128, 32). 
We apply mirror symmetry at the $z=0$ (equatorial) plane.
Before starting the calculation, we prepare six grid levels ($l=1-6$).
The base grid ($l=1$) has a box size of $L(1)=1.4\times10^5$\,au and a cell width of $h(1)=1.1\times10^3$\,au, respectively. 
The initial B.E. core is embedded in the fifth grid level ($l=5$). 
A uniform density of $n_{\rm ISM}=1.3\times10^5\,\cc$ is distributed outside the B.E. core. 
Once the calculation begins, a new finer grid is automatically generated to ensure the Jeans wavelength is resolved by at least eight cells. 
The finest grid is set to $l=11$, with a box size of $L(11)=137$\,au and a cell width of $h(11)=1.1$\,au.

\section{Results}
\label{sec:wholeevo}

\begin{figure*}
\begin{center}
\includegraphics[width=1.0\columnwidth]{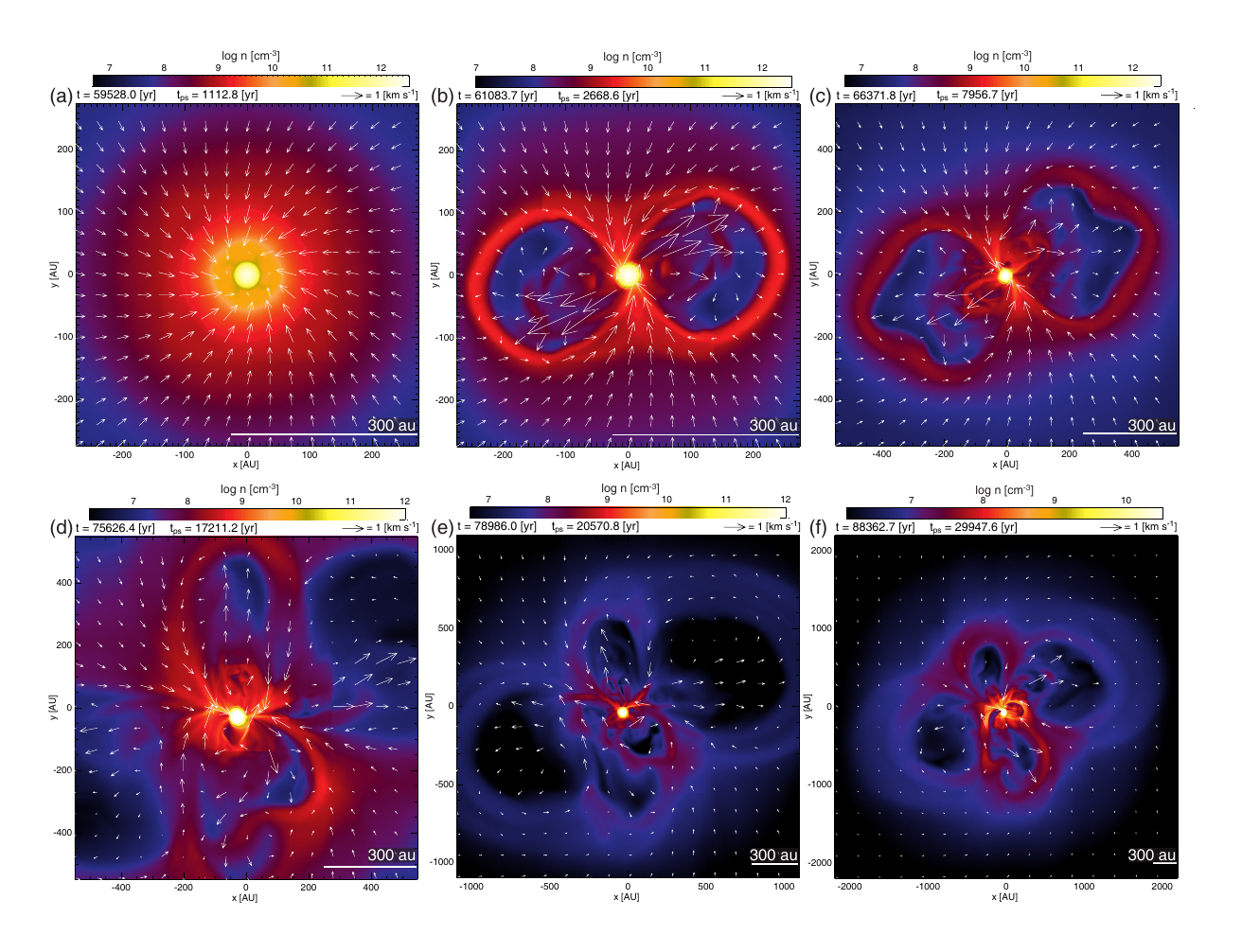}
\end{center}
\caption{
Time sequence of density (color) and velocity (arrows) distributions on the equatorial plane. 
The time $t$ after the calculation begins and the time $t_{\rm ps}$ after protostar formation are indicated in the top left corner of each panel. 
The spatial scale of each panel is different. 
For reference, a scale bar of 300 au is included in each panel.
An animated version of this figure is available. 
In the animation, the time sequence of the density on the equatorial plane (top left and bottom left panels) and on the $x=0$ plane (bottom right), as well as the magnetic field distribution (top right), as shown in Figures~\ref{fig:1}  and \ref{fig:2}, is displayed from the beginning to the end of the simulation.
The duration of the animation is 16\,s. (For a high-resolution animation see 
https://archive.iii.kyushu-u.ac.jp/public/L6JGgbAI9cskrW7zjQ2GGCD05syLw-gWRzdAAe90AEnm,  
https://vimeo.com/1041043487?share=copy)
}
\label{fig:1}
\end{figure*}

Once the central density of the collapsing cloud core reaches $\sim10^{10}\,\cc$, an adiabatic core (i.e., the first core, or protostar; \citealt{larson69}) forms at the center, as shown in Figure~\ref{fig:1}{\it a}. 
In this paper, we define the protostar formation epoch ($t_{\rm ps}=0$) as the time when the central or maximum density of the collapsing core reaches $10^{10}\,\cc$.

A bipolar cavity structure appears immediately after the formation of the protostar. 
Figure~\ref{fig:1}{\it b} shows that the cavity extends up to $\sim300$\,au from the protostar. 
The velocity within the cavity exceeds 2\,$\kms$ ($\sim$10 times the sound speed). 
The cavity structure extends to $\sim400$\,au at about 8,000\,yr after protostar formation (Fig.~\ref{fig:1}{\it c}). 
Thus, the expansion velocity of the cavity structure is $0.24\,\kms$, which roughly corresponds to the sound speed at $T=10$\,K.
The expansion velocity of the cavity is consistent with that estimated in observations \citep{tokuda23,tokuda24}. 

\begin{figure*}
\begin{center}
\includegraphics[width=1.0\columnwidth]{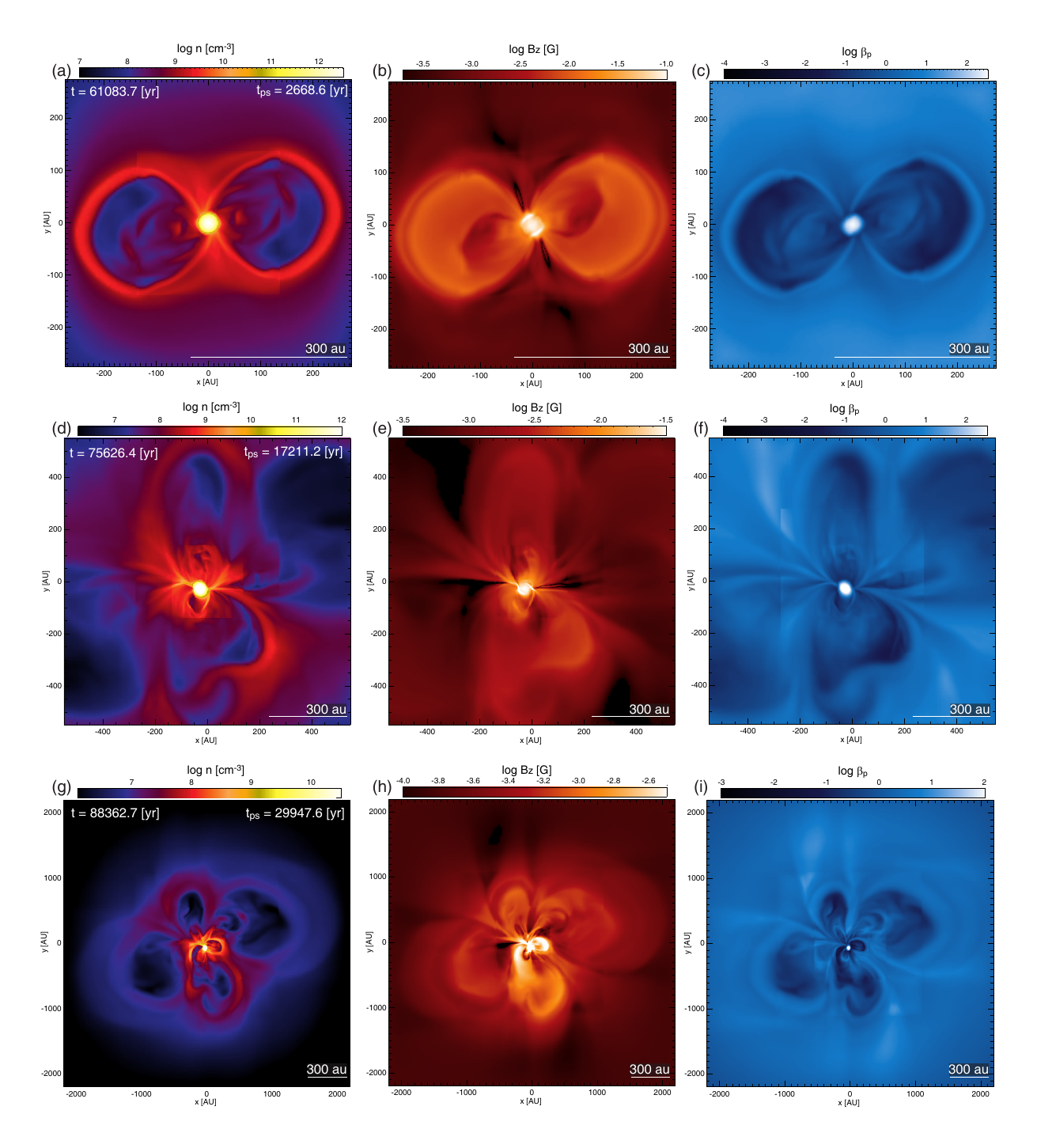}
\end{center}
\caption{
Density (left), $z$-component of the magnetic field (center), and plasma beta (right) distributions on the equatorial plane at $t_{\rm ps}=2,668.6$\,yr (top), 17,211.2\,yr (middle), and 29,947.6\,yr (bottom). 
The time $t$ after the calculation starts is also indicated in each panel of the left column. 
The spatial scale of the panels in each row is different.
For reference, a scale bar of 300 au is included in each panel.
It should be noted that, to produce each panel, we used physical quantities (gas density, magnetic field strength, and plasma beta) located closest to the $z=0$ plane, as there are no physical quantities exactly on each plane (or axis) in our numerical code \citep[see][]{Fukuda1999,Matsumoto2003a}.
Thus, the plotted quantities are slightly offset from the $z=0$ plane by $h(l)/2$, where $h(l)$ is the cell width of the $l$-th grid. The discontinuity seen at the grid boundaries (especially in Fig.~\ref{fig:2}{\it f} and {\it i}) is caused by the difference in the cell width between grids.
}
\label{fig:2}
\end{figure*}

The cavity structure is created by interchange instability. 
Figure~\ref{fig:2}{\it a}--{\it c}
shows the density (left), magnetic field strength (middle), and plasma beta (right) distributions on the equatorial plane at the same epoch as Figure~\ref{fig:1}{\it b}.
The magnetic field strength $B_z$ within the cavity is about $1-2$ orders of magnitude greater than that outside the cavity (Fig.~\ref{fig:2}{\it b}).
In addition, the plasma beta in the cavity is $\beta_{p}<10^{-2}$, while $\beta_{p}>1-10$ outside the cavity. 
This shows that magnetic flux is being expelled from near} the circumstellar disk, where there is a strong magnetic field.
We show that this is a robust effect that lasts throughout the main accretion phase of star formation ($> 10^4$ yr), going well past the 500 yr calculated by \citet{machida20b}.

\begin{figure*}
\begin{center}
\includegraphics[width=1.0\columnwidth]{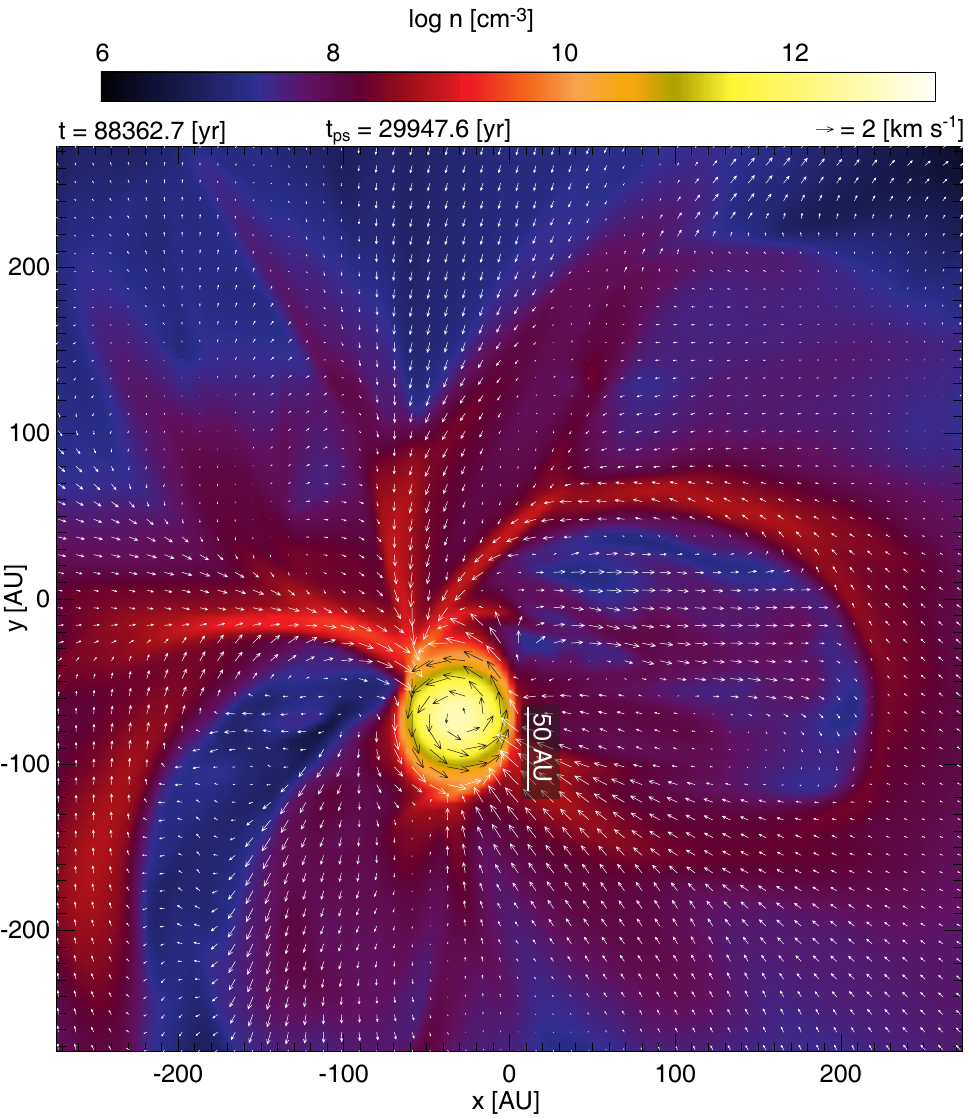}
\end{center}
\caption{
Density and velocity distribution around the circumstellar disk on the equatorial plane. 
The time $t$ after the calculation begins and the time $t_{\rm ps}$ after protostar formation are indicated at the top of the panel.
For reference, a scale bar of 50 au is included.
}
\label{fig:3}
\end{figure*}

In Figure~\ref{fig:1}{\it b} and {\it c}, we can see that new small cavities form within the larger cavities, indicating that interchange instability occurs recurrently.
This recurrence of the instability forms the complex structure shown in Figure~\ref{fig:1}{\it d}. 
The middle and bottom panels of Figure~\ref{fig:2} indicate that, although the density distribution is not uniform outside the disk, which corresponds to the yellow and white regions (see also Fig.~\ref{fig:3}), the low-density regions correspond to areas with a strong magnetic field and low plasma beta.

Figure~\ref{fig:1}{\it f} shows the density distribution about $3\times10^4$ yr after protostar formation and indicates that the disk is enclosed by several cavities. 
Figure~\ref{fig:3} is a close-up view of Figure~\ref{fig:1}{\it f}.
It suggests that a circumstellar disk with radius $\sim50$\,au 
forms and remains for at least $3\times10^4$ yr after protostar formation, even when interchange instability intermittently occurs.
In addition, Figure~\ref{fig:3} shows that mass accretion onto the disk occurs through channels between the cavities. 
It also shows that part of the accreting gas moves in a direction opposite to the disk (or protostar) because the gas is frozen to the magnetic field outside the disk, and the magnetic flux moves outward. 
At $\sim10^4$ yr the cavities extend up to $>1,000$\,au from the protostar (Fig.~\ref{fig:1}{\it f}). 
Figure~\ref{fig:2}{\it h} shows that the distribution of the magnetic field is highly complicated.
However, Figure~\ref{fig:2}{\it g}--{\it i} confirms that the low-density region is correlated with the strong magnetic field region. 
 
The structures formed by interchange instability exhibit similarities between observations and simulations: ring-like or horn-shaped structures are observed around the circumstellar disk. In addition, the sizes of the structures are comparable between the simulations and observations.
Furthermore, the expansion velocity of the observed ring structures is approximately the sound speed, which is consistent with the simulation results.

\begin{figure*}
\begin{center}
\includegraphics[width=1.0\columnwidth]{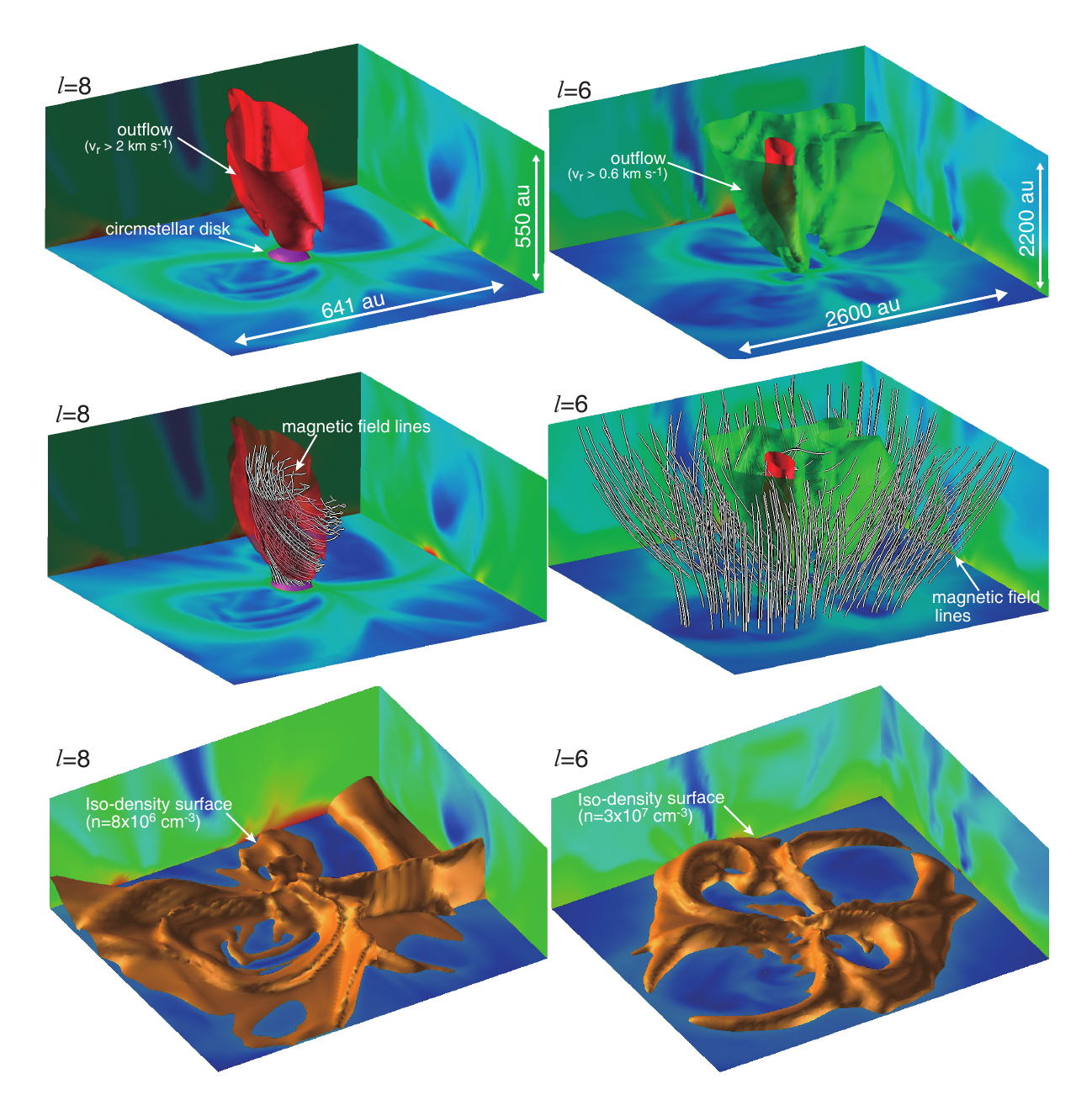}
\end{center}
\caption{
Three-dimensional structure of the outflow (red and green surfaces), magnetic field (black and white lines) and high-density region (orange surface) at $t_{\rm ps}=29,947.6$\,yr with $l=8$ (left) and $l=6$ (right) grids. 
The middle panels are the same as the top panels, but with magnetic field lines plotted. 
The high-density region of $n=8\times10^6\,\cc$ (bottom left) and $3\times10^7\,\cc$ (bottom right) are plotted in the bottom panels.  
The density distribution on the $z=0$, $x=0$, and $y=0$ planes is projected onto each wall surface. 
The circumstellar disk is represented by the purple surface, where the density $n >10^{10}\,\cc$.  
An animated version of this figure is available. 
In the animation, the time sequence of the outflow ($v_r>2\,\kms$) as in Figure~4 top right panel ($l=6$) from the beginning until the end of the simulation is shown. The duration of the animation is 19\,s. (For a high-resolution animation see
https://archive.iii.kyushu-u.ac.jp/public/L6JGgbAI9cskrW7zjQ2GGCD05syLw-gWRzdAAe90AEnm, https://vimeo.com/1041043593?share=copy)
(An animation of this figure is available in the online article.)
}
\label{fig:4}
\end{figure*}

\citet{machida20b} showed that a weak magnetically-driven outflow appears even as interchange instability occurs. 
However, the evolution of the outflow was calculated for only $\sim500$ yr after protostar formation. 
Thus, it was not clear whether the outflow could be sustained for a long duration.
Figure~\ref{fig:4} shows a three-dimensional view of the outflow, magnetic field lines and density structure on a small scale (left) and a large scale (right) at the same epoch as Figure~\ref{fig:1}{\it f} ($t_{\rm ps}=29,947.6$\,yr). 
The top panels of Figure~\ref{fig:4} indicate that the outflow is driven from near the edge of the circumstellar disk (see \citealt{Basu24} and \citealt{machida24} for details of the wind driving in less magnetized models).
In addition, the magnetic field lines are strongly twisted in the outflow (Fig.~\ref{fig:4} middle left panel). 
On a larger scale (Fig.~\ref{fig:4} top right), the high-velocity component of the outflow is enclosed by the low-velocity component. 
In the middle right panel of Figure~\ref{fig:4}, the magnetic field lines followed outward from the density cavity are slightly inclined but remain mostly straight without being strongly twisted. 
Although the outflow is driven by the circumstellar disk with a density $\gtrsim 10^{10}\,\cc$, the bottom panels of Figure~\ref{fig:4} show that multiple rings and cavities surround the circumstellar disk.

\begin{figure*}
\begin{center}
\includegraphics[width=1.0\columnwidth]{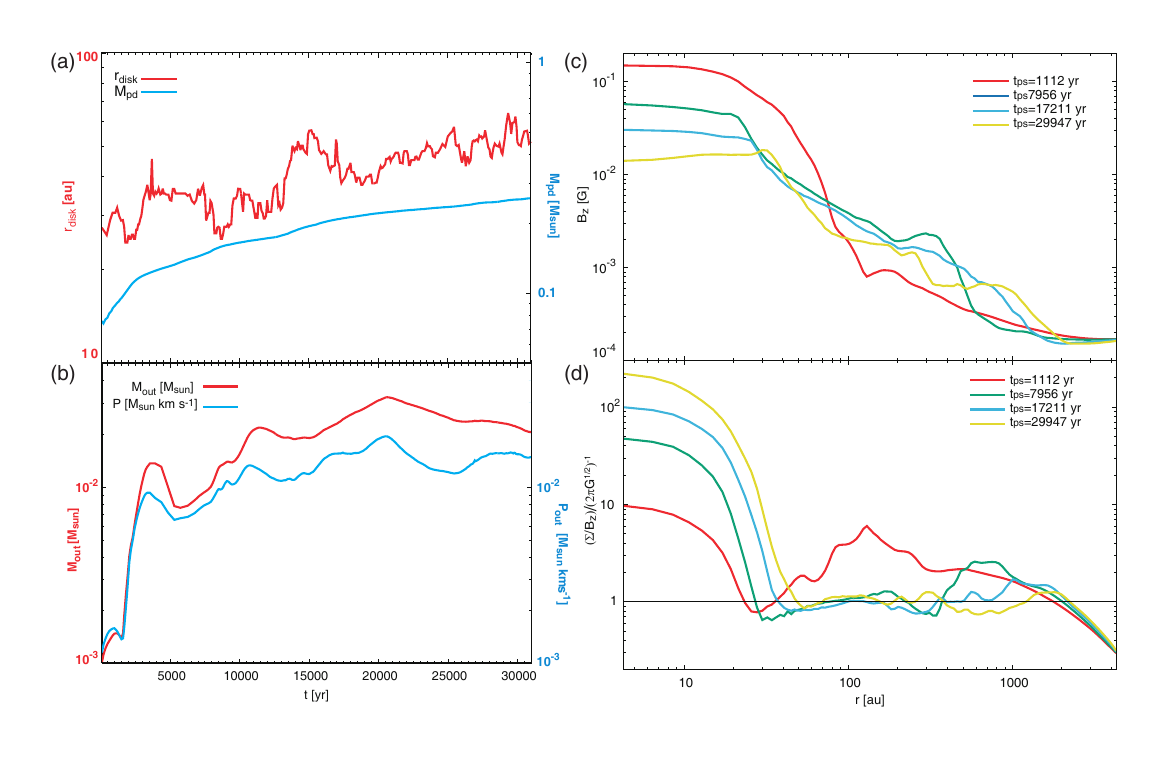}
\end{center}
\caption{
(a) Disk radius (red; left axis) and mass (blue; right axis) as a function of time after protostar formation.  
(b) Outflow mass (red; left axis) and outflow momentum (blue; right axis) as a function of time after protostar formation.  
(c) Azimuthally averaged vertical magnetic field component, $B_z$, as a function of distance from the protostar at four different epochs.
(d) Azimuthally averaged local mass-to-flux ratio, $\Sigma / B_z$, normalized by its critical value, $(2\pi G^{1/2})^{-1}$, as a function of distance from the protostar at four different epochs.
}
\label{fig:5}
\end{figure*}

To quantitatively investigate the properties of the disk and outflow, the disk radius and mass are plotted as a function of time after protostar formation in Figure~\ref{fig:5}{\it a}. 
To identify the disk, we determine the location with the maximum density as the position of the protostar (or the disk center).  
We calculate the radial $v_{r}$ and azimuthal $v_{\phi}$ velocity components in cylindrical coordinates with respect to the position of the protostar.
Then, we estimate the disk's minimum density, $\rho_{\rm d}$. 
We define the minimum density of the disk as the lowest density where $v_{\phi} > f_{\rm disk}\, v_{r}$ is satisfied on the equatorial plane, in which $f_{\rm disk} = 5$ was adopted after trial and error.
The disk mass (plus protostellar mass) is estimated as 
\begin{equation}
M_{\rm pd} = \int_{\rho > \rho_{\rm d}} \rho \, dV.
\end{equation}
We define the disk radius as the distance to the farthest point from the central star where $v_{\phi} > f_{\rm disk} v_{r}$ is satisfied. 
Figure~\ref{fig:5}{\it a} shows that although there is variability in the time evolution, the disk radius increases from $\sim25$\,au to $\sim50$\,au over $3\times10^4$ yr. 
The mass of the disk and protostar reaches $M_{\rm pd}=0.27\,\msun$ at the end of the simulation.  
Thus, the disk continues to grow gradually even as interchange instability occurs around the edge of the disk.

The mass and momentum of the outflow are plotted as a function of time after protostar formation in Figure~\ref{fig:5}{\it b}. 
Firstly, we define the outflow as the region where the gas has a $z$-direction velocity greater than the isothermal sound speed, $v_{z} > c_{s}$ for $z > 0$ ($v_z < -c_{s,0}$ for $z < 0$), where $c_{s,0} = 0.2,\kms$, corresponding to an isothermal temperature of 10\,K, and $v_z$ is the vertical velocity component in cylindrical coordinates.
We use the $z$-component of the velocity to exclude the gas expansion caused by interchange instability (for details, see below). 
As shown in Figure~\ref{fig:1}, the gas expansion velocity in the cavities exceeds the sound speed.
Figure~\ref{fig:5}{\it b} shows that the outflow mass reaches $M_{\rm out} \gtrsim 0.01\,\msun$ by the end of the simulation. 
This is about 10\% of the infalling gas, so that the outflow mass is about 10\% of the disk and protostellar mass. 
The outflow momentum is $P_{\rm out}\sim 10^{-2}\,\msun\,\kms$ in the time range $5\times10^3\, {\rm yr} \lesssim t_{\rm ps} \lesssim 3\times10^4\, {\rm yr}$, which is comparable to observations  (\citealt{machida13} and references therein).  
Therefore, we find that even when interchange instability occurs and forms a complex envelope around the disk, the outflow does not 
weaken significantly. 
Note that the outflow in the early accretion phase ($t_{\rm ps}\lesssim 1,000$\,yr) is very weak, which is consistent with \citet{machida20b}. 
However, the outflow becomes powerful for $t_{\rm ps} \gtrsim 1,000$\,yr.

Next, to estimate the total mass inflowing into and outflowing from the star and disk system, we examine the inflow and outflow rates, taking into account both the outgoing flows caused by interchange instability and the disk wind (or outflow).
We estimated the mass inflow and outflow rates on the $l=7$ and $l=8$ grid surfaces, which have box sizes of $L(l=7) = 2,200$\,au and $L(l=8) = 1,100$\,au, respectively \citep{machida24}.
The mass inflow rate is calculated as
\begin{equation} 
\dot{M}_{\rm in}(l=7, 8) = \int_{\rm surface \, of \, {\it l}=7, 8 \, \rm grid} \rho\, \vect{v} \cdot \vect{n} (<0)\, dS, 
\end{equation} 
where $\vect{n}$ is the normal vector of each surface, and the integration is performed only when  $\vect{v} \cdot \vect{n} < 0$ (i.e., inflow).
The mass outflow rate consists of two components: the outgoing flow due to the disk wind ($\dot{M}_{\rm out, DW}$) and the outgoing flow caused by interchange instability ($\dot{M}_{\rm out, IC}$).
We confirmed that gas flows out laterally due to interchange instability when $v_r > v_z$.
On the other hand, the outgoing flow originating from the disk wind (or protostellar outflow) satisfies the condition $v_z > v_r$.
The outflow rate due to the disk wind is calculated as
\begin{equation} 
\dot{M}_{\rm out, DW}(l=7, 8) = \int_{\rm surface \, of \,  {\it l}=7, 8 \, \, grid} \rho \, \vect{v} \cdot \vect{n} (>0) \, dS, \quad {\rm for} \ v_z > v_r, 
\label{eq:mDW}
\end{equation}
and the outflow rate due to interchange instability is calculated as
\begin{equation} 
\dot{M}_{\rm out, IC}(l=7, 8) = \int_{\rm surface \, of \, {\it l}=7, 8  \, grid} \rho \, \vect{v} \cdot \vect{n} (>0) \, dS, \quad {\rm for} \ v_r > v_z. 
\label{eq:mIC}
\end{equation}
For equations~(\ref{eq:mDW}) and (\ref{eq:mIC}), the integration is performed when  $\vect{v} \cdot \vect{n} > 0$ (i.e., outgoing flow).

\begin{figure*}
\begin{center}
\includegraphics[width=0.7\columnwidth]{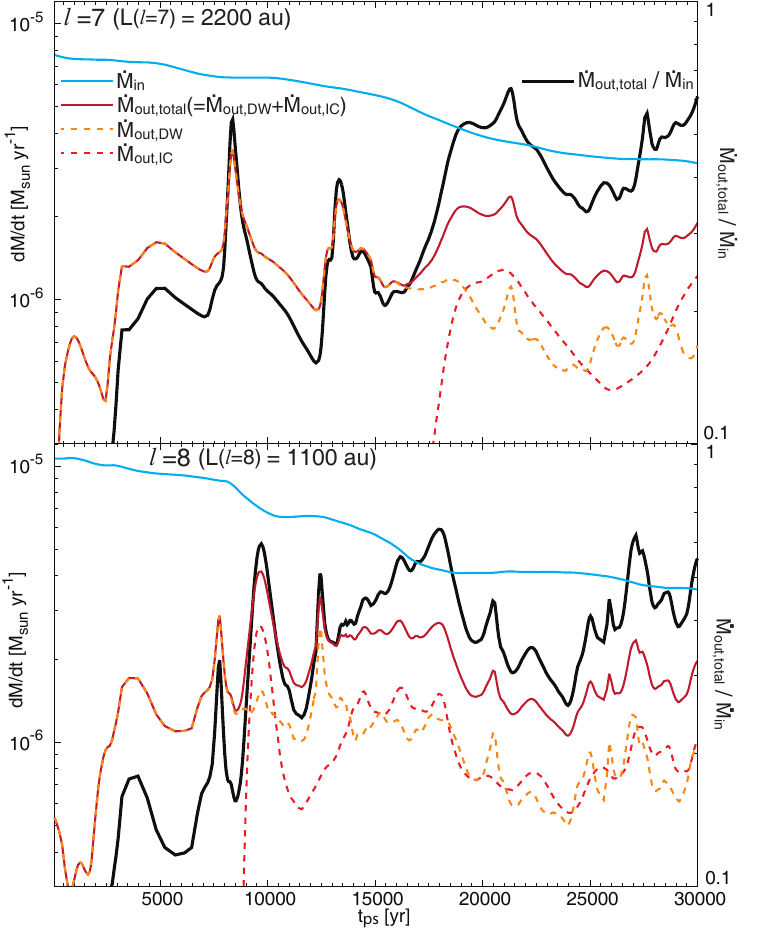}
\end{center}
\caption{
Mass inflow and outflow rates (left axis) for box scales of 2,200\,au ($l=7$, top) and 1,100\,au ($l=8$, bottom) are plotted against the elapsed time after protostar formation. The total outflow rate is composed of contributions from the disk wind (or protostellar outflow) and interchange instability. The ratio of inflow to outflow (right axis) is also plotted against the elapsed time after protostar formation.
}
\label{fig:6}
\end{figure*}

Figure~\ref{fig:6} shows the mass infall and outflow rates, with the mass outflow rates due to the disk wind, interchange instability, and the total outflow rate plotted.
The figure indicates that the mass inflow rate gradually decreases from $\sim 10^{-5}$ to $\sim 4\times 10^{-6}\,\msunyr$ for both $l=7$ and 8 grids.
Both the mass outflow rates due to the disk wind and interchange instability oscillate, indicating that mass ejections caused by the disk wind and interchange instability are time variable.
In the early phase, only the disk wind contributes to the mass outflow rate at both the 1,100 au and 2,200 au scales. 
For $t_{\rm ps}\gtrsim (1-2)\times10^4$\,yr, gas flows caused by interchange instability also contribute to the mass ejection.
The mass outflow rate due to interchange instability becomes comparable to that of the disk wind in the later phase.
In addition, the ratio of the infall to outflow rates is in the range $\dot{M}_{\rm out, total}/\dot{M}_{\rm in} \sim 0.3-0.6$.
Thus, while further time integration is necessary, the mass ejection caused by interchange instability may play a significant role in the early stages of star formation.

To examine the magnetic field distribution, the component $B_z$ is plotted against the distance from the protostar in Figure~\ref{fig:5}{\it c}, where $B_z$ is averaged azimuthally.
The figure shows that the magnetic field strength near the center ($\lesssim20$\,au) is about 0.1\,G at approximately 1,000 yr after protostar formation. 
As seen in Figure~\ref{fig:5}{\it c}, the field strength around the center gradually decreases over time. 
By the end of the simulation, the central field strength is approximately 0.01\,G. 
Thus, the magnetic field near the center weakens by roughly one order of magnitude over $3\times10^4$ yr. 
In contrast, the magnetic field strength in the outer region ($r\gtrsim20$\,au) increases over time. 
Around $r=1,000$\,au, the magnetic field at $t_{\rm ps}=29,947$\,yr is stronger than at other times. 
As shown in Figure~\ref{fig:1}, the cavity created by the interchange instability extends beyond 1,000\,au at $t_{\rm ps}\sim3\times10^4$ yr.

\begin{figure*}
\begin{center}
\includegraphics[width=0.8\columnwidth]{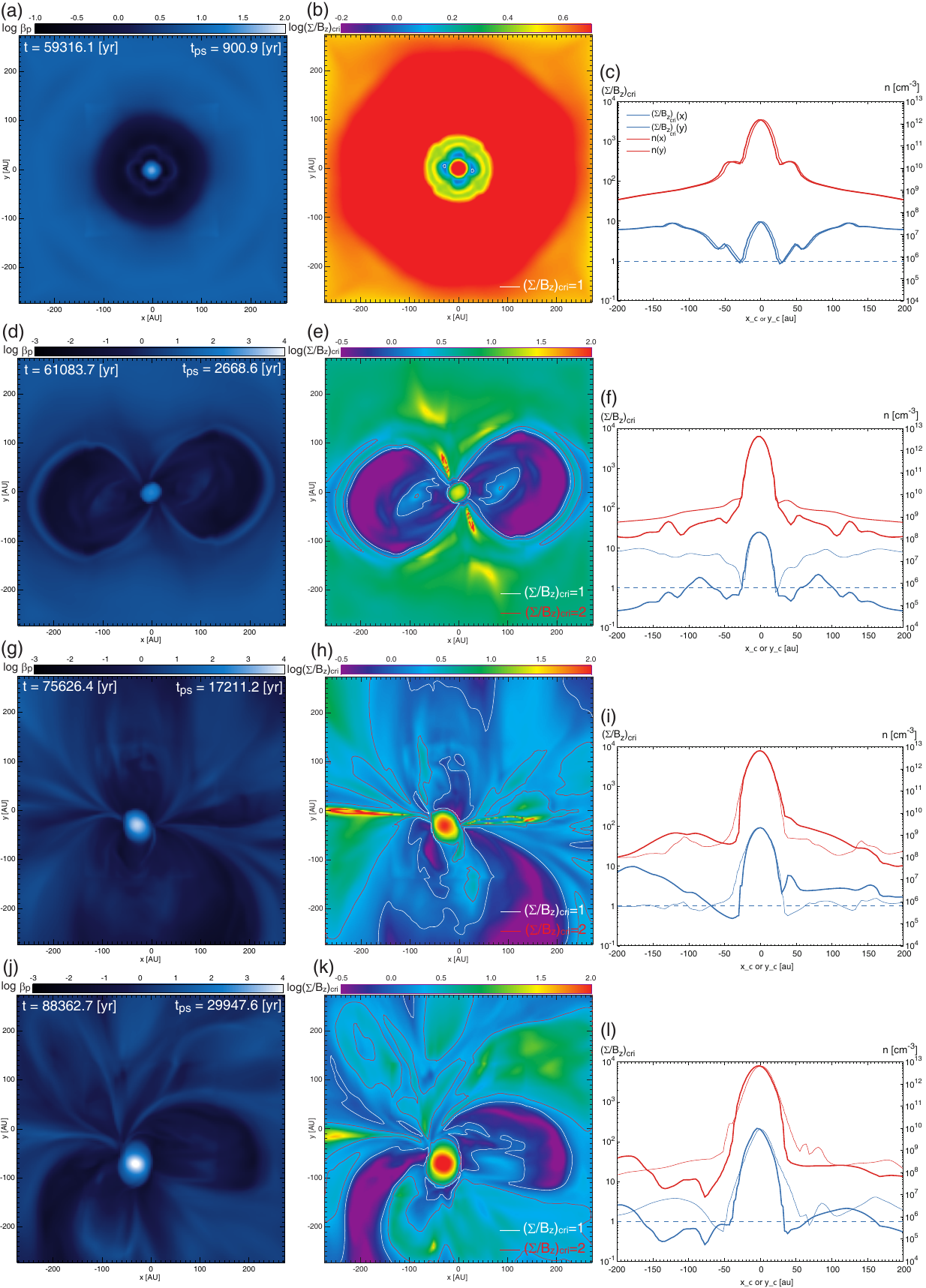}
\end{center}
\caption{
(Left panels) Plasma beta, $\beta_{\rm p}$, on the equatorial plane.
The color bar in panel ({\it a}) is different from those in panels ({\it d}), ({\it g}), and ({\it j}).
(Middle panels) The ratio of surface density to the $z$-component of magnetic field strength normalized by the critical value, $(\Sigma/B_z)_{\rm cri}$, on the equatorial plane. 
The contours of $(\Sigma/B_z)_{\rm cri}=1$ (white) and $2$ (red) are plotted in each panel.
The color bar in panel ({\it b}) is different from those in panels ({\it e}), ({\it h}), and ({\it k}).
(Right panels) Density ($n(x)$ and $n(y)$; right axis) and the ratio of surface density to the magnetic field strength ($(\Sigma/B_z)_{\rm cri}(x)$ and $(\Sigma/B_z)_{\rm cri}(y)$; left axis) along the $x$- and $y$-axes  on the equatorial plane.
The physical quantities are plotted at four different epochs. 
}
\label{fig:7}
\end{figure*}

To confirm the importance of the magnetic field, the plasma beta, $\beta_{\rm p}$, on the equatorial plane around the central region is plotted in the left panels of Figure~\ref{fig:7}.
The plasma beta within the disk at each epoch reaches $\beta_{\rm p} \gtrsim 10^2 - 10^4$, whereas $\beta_{\rm p} \ll 1$ outside the disk, as described above. 
Thus, the magnetic field plays a significant role outside the disk, as also seen in Figure~\ref{fig:2}.

Figure~\ref{fig:5}{\it d} shows the ratio of the surface density, $\Sigma$, to the vertical component of the magnetic field, $B_z$, normalized by its critical value, $(2\pi G^{1/2})^{-1}$, as a function of the distance from the protostar, where the ratio is azimuthally averaged.
Note that the surface density and scale height of the disk or outer disk-like region (pseudodisk) are taken to be $\Sigma=\rho h$ and $h=c_s/(\pi G \rho)^{1/2}$, respectively \citep{scott80,saigo98}.
There is an inversion (outward gradient) of the mass-to-flux ratio outside the disk radius of $30-50$ au at various locations and times. 
These gradients are then reduced at later times through the action of the interchange instability, which dynamically moves magnetic flux outward and brings the local mass-to-flux ratio back toward the critical value.

To examine the ratio in detail, we plotted $(\Sigma/B_z)_{\rm cri}$ on the equatorial plane at different epochs in the middle panels of Figure~\ref{fig:7}. 
In addition, the ratio along the $x$-axis, $(\Sigma/B_z)_{\rm cri}(x)$, and $y$-axis, $(\Sigma/B_z)_{\rm cri}(y)$, are plotted in the right panels of Figure~\ref{fig:7}.
Figure~\ref{fig:7}{\it b} and {\it c} show that $(\Sigma/B_z)_{\rm cri}$ has a local minimum just outside the disk ($r\sim 30$\,au). 
In addition, $(\Sigma/B_z)_{\rm cri}(x)$ is slightly lower than $(\Sigma/B_z)_{\rm cri}(y)$ outside the disk.  
Then, interchange instability occurs in the $x$-direction and creates a cavity structure, as shown in Figure~\ref{fig:7}{\it e} and {\it f}. 
We can confirm a complicated structure created by interchange instability outside the disk in Figure~\ref{fig:7}{\it e}, {\it h}, and {\it k}. 
After interchange instability occurs, the continuing evolution causes the ratio $(\Sigma/B_z)_{\rm cri}$ to again become lower than unity near the disk.  
There is once again a local minimum of $(\Sigma/B_z)_{\rm cri}$ just outside the disk in Figure~\ref{fig:7}{\it f}, {\it i}, and {\it l}. 
This indicates that interchange instability occurs repeatedly after the first instability occurs. 

\section{Discussion and Conclusions}
Finally, after discussing the spatial resolution necessary to investigate interchange instability and the evolutionary stage considered in this study, we conclude our results.

The spatial resolution in this study differs from that in our previous work.
The minimum resolution in this study is $h(l_{\rm max}) = 1.1$\,au (see, \S\ref{sec:model}), which is coarser than the $h(l_{\rm max}) = 0.01$\,au resolution adopted in \citet{machida20b}. 
Despite the lower resolution, we have sufficiently resolved the first core, which has a size of $\gtrsim 10$\,au, where magnetic dissipation occurs \citep{machida07,machida18}.
The first core directly evolves into a rotationally supported disk or circumstellar disk \citep{machida12}, and interchange instability occurs at the edge of the first core or circumstellar disk, where magnetic flux accumulates \citep{machida20b}.
Therefore, it is possible to investigate interchange instability as long as the first core or circumstellar disk is adequately resolved.
However, differences in resolution can lead to subtle variations in the density and magnetic field distributions between low- and high-resolution simulations.
As a result, properties of the interchange instability, such as the timing of its onset and the growth of cavity structures, may differ quantitatively.
While these aspects should be addressed in future studies, we believe this study sufficiently captures the structures formed by interchange instability outside the first core or circumstellar disk, provided the outer edge of these regions is resolved.

As described in \S\ref{sec:model}, interchange instability can occur numerically when a sink cell is introduced.
To avoid this issue, we adopted an adiabatic EOS method.
We consider that the adiabatic core appearing in the simulation represents both the protostar and the circumstellar disk.
However, its size is larger than that of the protostar because part of the core is supported by thermal pressure, similar to the first (Larson) core \citep{larson69}.
Based on core collapse simulations \citep{Tomida2010,tomida13}, the first (or adiabatic) core can persist for $\sim10^4$\,yr before it begins to shrink significantly.
After this point, the protostellar (or adiabatic core) radius is expected to become smaller than the maximum resolution of the simulation.
Therefore, it can also be interpreted that the simulations in this study cover the earliest phases of protostellar evolution, during which the adiabatic (Larson) core remains resolved.
Although the long-term evolution beyond this point is not covered in this study, we believe the current approach adequately captures the dynamics of interchange instability during the early stages.

We investigated the evolution of a cloud core with a strong magnetic field and performed the simulation until about $3\times10^4$ yr after protostar formation. 
In the simulation, we confirmed many features in the disk and circumstellar region that are associated with interchange instability.  
The simulation showed that interchange instability and a significant removal of the magnetic flux from the disk recurrently occurs during the mass accretion phase.
The structures formed by interchange instability such as arcs, cavities, and spikes, are maintained for $>10^4$ yr and extend to distances $>1,000$\,au.
We found that even with the significant removal of magnetic flux from near the disk edge, the disk continues to grow gradually, although its size remains small. 
An outflow is initially weak in the strongly magnetized cloud, but strengthens considerably after $\sim 10^3$ yr following protostar formation.
Our findings explain the complex structures around circumstellar disks seen in recent observations  \citep{tokuda23,tokuda24} and strongly imply that the interchange instability is occurring there. 

The interchange instability is closely tied to the disk formation process. Thus, its form and strength, or whether it occurs at all, is influenced by many factors. These include the magnetic field strength, rotation rate, and thermal stability of the prestellar core \citep[see review by][]{tsukamoto23b}. The effects of the EOS and ohmic and ambipolar resistivities are also crucial. The stiff EOS we have used creates a strong pressure gradient at densities $> 10^{10}\,\cc$, and other implementations of the EOS may lead to a different interplay with the magnetic pressure gradient, which needs to exceed the pressure gradient for the interchange instability to occur \citep{spruit90,lubow95}. Future work can explore a wide parameter space of initial conditions and microphysics, while maintaining high spatial resolution and long time integration.

\section*{Acknowledgements}
This research used the computational resources of the HPCI system provided by the Cyber Science Center at Tohoku University and the Cybermedia Center at Osaka University (Project ID: hp230035, hp240010).
Simulations reported in this paper were also performed by 2023, 2024 Koubo Kadai on Earth Simulator (NEC SX-ACE) at JAMSTEC. 
The present study was supported by JSPS KAKENHI Grant (JP21H00046, JP21K03617: MNM).
This work was supported by a NAOJ ALMA Scientific Research grant (No. 2022-22B). 
S.B. was supported by a Discovery Grant from NSERC.

\bibliography{machida}{}
\bibliographystyle{aasjournal}

\end{document}